# A new approach for scientific data dissemination in developing countries : a case of Indonesia

L.T. Handoko

handoko@teori.fisika.lipi.go.id

*Group for Theoretical and Computational Physics, Research Center for Physics, Indonesian Institute of Sciences*
*Kompleks Puspiptek Serpong, Tangerang 15310, Indonesia*
*http://teori.fisika.lipi.go.id*

*Abstract*

This short paper is intended as an additional progress report to share our experiences in Indonesia on collecting, integrating and disseminating both global and local scientific data across the country through the web technology. Our recent efforts are exerted on improving the local public access to global scientific data, and on the other hand encouraging the local scientific data to be more accessible for the global communities. We have maintained well-connected infrastructure and some web-based information management systems to realize such objectives. This paper is especially focused on introducing the ARSIP for mirroring global as well as sharing local scientific data, and the newly developed Indonesian Scientific Index for integrating local scientific data through an automated intelligent indexing system.

*Keyword(s) : information management, mirror service, scientific index*

## 1. INTRODUCTION

Nowadays, in the (especially digital) information age, information is becoming the most important asset in any aspects of human life. Also in scientific fields, researchers and academicians across the globe are faced in very tough competition. In the last decades, the internet technology changed the way of information sharing to be much faster and more accurate than before. Everything is available in almost real time which then leads to prompt knowledge upgrading and enrichment. This fact is reflected in the number of scientific publications which is increasing exponentially [1].

Such revolutionary waves on digital information influences all regions, including the developing countries like Indonesia, although the main sources for in particular scientific information are still dominated by our partners in the developed regions. Most of us in the developing countries still behave as hungry consumers, and suffer from the domestic problems to take a primary role as information providers in the global communities. On the other hand, no one doubts that even developing countries have rich and valuable scientific information which should be contributed to the human civilization.

In regard to the scientific data dissemination, there are several considerable obstacles which could probably keep us away from the dense traffic of global scientific information :
1. Lack of infrastructure :
    Along with internet technology, the limited local gateway to the international backbone could lead to a severe situation for local scientific communities to access global scientific information that is very crucial to compete equally with their counterpart in the rest of the

world.
2. Poor human resources :
    This is the main key to realize smooth and smart access to global data and also share our own data to the others.
3. Less scientific activities :
    This is a fundamental problem in most developing countries with short scientific culture. However, this might be improved subsequently if the first and second points above are improved properly. This leads also to low awareness on well-managed scientific information.

Now we are going to explain in detail the solutions we have deployed so far.

The first problem, the lack of infrastructure, is very hard to resolve since it involves many parties, many outside the scientific communities. We are compelled to compromise with the macro economy of our country. Because, connecting to the international gateway with sufficient backbone leads to nontrivial financial consequences for the local internet service providers. In the case of Indonesia, we fortunately enjoy more bandwidth with much reduced cost after privatization of telecommunication sector recent years.

However, we in Indonesia still suffer from poor human resources. More accurately, we are not yet provided with enough funding to attract potential local human resources around us to develop management information systems in, as many as possible, local scientific institutions. This problem is also closely related to low awareness of the importance of good information management among the related institutions. We should admit that top managements in most scientific institutions still lack awareness of this matter.

The last, but may be the most important, one is weak and small scale of scientific activities itself. Of course, no one expects relevant scientific outputs if there is no scientific activity at all. In contrast to India and China, we have very short history and poor culture on scientific activities. Although significant advances are already on the track in recent years, we still need time, perhaps a few generations to place it on a solid footing. However, rapid information exchange through internet could also boost the improvement of local scientific activities.

In this paper, we explain the solution to deal with the second and last problems mentioned above. The presentation is focused on the real solutions we have developed so far in Indonesia. Because in our point of view, the first problem is getting resolved in Indonesia, while the rest should still be improved. More importantly the remaining problems involve more the scientific communities itself. Therefore it should be resolved by ourself, or at least we have to take a leading role in that. In addition to the mirroring services, we also introduce the newly developed Indonesian Scientific Index as a key-solution to integrate local scientific data easily.

## 2. THE APPROACH : ACCESSING THE GLOBAL AND SHARING THE LOCAL

Our strategy is based on an underlying assumption that the internet is already widely used by at least local scientific communities in Indonesia who are our main targeted market. Because according to the study by the Internet World Stats, Asia including Indonesia enjoys an average use growth of 300% annually [2].

In spite of low computer penetration in Indonesia, even compared with the surrounding South East Asia countries, the Indonesians sit on the most active internet users in the world. For instance, the Indonesian is the seventh most active blogger. So, the low computer penetration might more be caused by our huge population and wide but separated lands..More precisely, referring to the statistic by the Indonesian Ministry of National Education, most of higher education institutes and Research Centers are already connected by internet [3]. Moreover, since two years ago, thousands of high schools in the cities are each provided with a computer laboratory with internet permanent access. No need to say about internet cafes spread across the country. Concerning these facts, we could assume that the majority of academicians and researchers in Indonesia presently have internet access at a reasonable cost.

Under the assumption mentioned above, we have developed some systems using the internet technology as the easiest and appropriate way to improve local accessibilities to the global data as well as to share the local data globally. Our strategy goes to two approaches :

1. Accessing the global data :
   We pay a great effort to improve the accessibility of local users to the global scientific data. This is important to fill the information gap between local researchers and their partners abroad. The service is crucial especially for scientific communities required huge raw data like biotechnologist with the Protein Data Bank, etc.
2. Sharing the local data :
   In order to enhance the existence of local researchers and their activities to the global communities should play an important role. Enabling global access could lead to a new kind of collaboration since it provides a great modality for local people to collaborate with global counterparts as equal partners.

Below we give more detail description on both approaches.

Actually mirroring services are commonly known since the internet age in the beginning of the 90's. This approach is mainly motivated by a need to reduce the interconnection cost of international bandwidth if everyone goes directly to the main sources abroad. The service would isolate local access within the local backbone, and subsequently speed up the data retrieval. However, reversing such services to host local scientific data for global communities is a new approach. This approach is a breakthrough to solve variable and mostly limited bandwidth in most institutions across the country. Normally local backbone is free of charge and almost unlimited, while the international gateway is limited according to the allocated bandwidth subscribed by the national internet providers. One can imagine easily, this could cause a bottleneck in case of simultaneous huge connections from both local and global sides. This leads to severe delays to either local repository from abroad or global repository from local.

We resolve this endless problem by inviting the main Indonesian internet providers having direct international backbones. After lengthy negotiations, the Indosat Mega Media (IM2) [4] agreed to support our project under an official Memorandum of Understanding with the Indonesian Institute of Sciences (LIPI) in 2003. The IM2 provides unlimited access to our servers which are collocated in their main Network Operation Center (NOC). Since the NOC is directly connected to the international backbone and the local Indonesian Internet Exchange (IIX) [5] as well, the connection problem from global and local users has been resolved completely. Furthermore, the upstream connection to the local data would not disturb the limited bandwidth subscribed by the institutions own the data.

Now we are going to introduce what we have really done so far in the subsequent sections. First we explain the common mirroring service, the so-called ARSIP, followed by the Indonesian Scientific Index as a national scale attempt to integrate local scientific data.

## 3. ARSIP : MIRROR SERVER FOR SCIENTIFIC DATA

Arsip in Indonesian language means archive. ARSIP provides a mirroring service for global scientific data since 2003 [6]. It follows completely the mainstream in the global scientific communities by mirroring major and globally recognized scientific data in our servers.

The main difference to other mirroring services is that ARSIP mirrors a wide range of scientific data, from physics to information technology, no matter what the field. The reason is because mirroring such global scientific data requires sophisticated hardware and infrastructure like huge repository space which should be upgraded occasionally to anticipate the data growth and of course unlimited bandwidth to enable simultaneous access and routine data synchronization. As mentioned earlier, very few institutions have such capabilities beside the fundamental problem of limited bandwidth. By centralizing the infrastructure and supporting human resources, we can provide a world class mirroring service efficiently.

We should remark that the physical hardware and infrastructure are actually not enough to realize a mirroring service. It also requires lengthy negotiation with the original data owner. In most cases it is much easier to pass it to some colleagues who already have close relationship with them anyway. This is the reason that the ARSIP team consists of many active researchers in the wide range of research fields. Therefore, despite centralized hardware the negotiation processes should be decentralized and distributed to, as many as possible, appropriate people in the related fields. Again, this approach is rather new and has succeeded in mirroring some global scientific data especially in the field of physics so far. For instance, we have maintained official mirrors for the Astrophysics Data System maintained by NASA [7], the Particle Data Group owned by LBL [8], the High Energy Physics Spires run by SLAC at Stanford University [9] and so on.

As already pointed out previously, ARSIP also mirrors local scientific data to improve international access by potential global partners. In the case of Indonesia, we have very rich scientific data related to our biodiversity, rare natural phenomena due to our geographical location and so forth. Through the ARSIP mirroring services, the local institutions maintain and generate the scientific data, and share the data with their counterparts more reliably due to guaranteed bandwidth and uptime. They just need to synchronize the mirrored data to ARSIP on a periodic basis through local backbone like IIX. This would clearly improve the reliability of global access to the local sources without scarifying the limited bandwidth of each institution. There is also another unfathomable advantage for the data owners, that is their data in ARSIP might be used as the backup in some terrible cases causing data losses etc. This could again reduce the overall maintenance cost at the institutions.

We have started this service recently, in the beginning of 2007. For instance, we are now in the process mirroring the earth magnetic observation collected by the Indonesian National Space Agency (LAPAN). The project is a global collaboration initiated by the SERC at Kyushu University [10]. Another recent attempt is to mirror the astronomy observation data by the Bosscha Observatory [11]. This kind of data perfectly fits our objectives since the data are mostly images at huge size. If the global users directly downloaded the data from the original source, it would exhaust the total bandwidth of the Bandung Institute of Technology.

## 4. ISI : INDONESIAN SCIENTIFIC INDEX

Now let us introduce briefly our newly developed approach to integrate the local scientific data more accurately and efficiently. Following the mirroring services we have implemented, now we deal with another problem on collecting the local scientific data. This is motivated by (local) public needs to access scientific data generated by local institutions as a part of public transparency and responsibility.

Although some institutions have already published and made their scientific output open to the public, it is hard to collect all of them consistently and keep ourself up-to-date. Actually, since 2005 we have implemented a national scale scientific information management system, namely DBRiptek maintained by the Indonesian Ministry of Research and Technology[12]. The system allows multiple representatives from all scientific institutions including universities to login and maintain their own scientific data. However, this approach depends highly on the integrity and consistency of each representative. Moreover, it wastes valuable time and human resources since the data entry has to be done twice : first for their own websites and the second for DBRiptek.

Concerning these problems we then proposed a new approach inspired by the crawling technology pioneered by Google [13]. However, we have modified the basic work-flow to fit our special objectives. In addition to the basic automated crawl, at the back-end we have configured manually each targeted sources to enable well-structured index databases. So, each targeted website has unique initial configuration. Using well-structured index databases we are able to provide much better search results to meet users' expectations. For instance, our system could distinguish if the crawled pages should belong to the publication, personal profile, bibliography, institution profile and so on. Also it could separate the title from abstract from a publication web page.

Deploying this approach, we just require the participating institutions to open their systems to our crawling machine, and of course make all shared information accessible by the public through the web. During the present running test, we are happy that the system works well as expected. We then named as the Indonesian Scientific Index – ISI [14].

Through ISI, we are able to "integrate" the local scientific data in an efficient (timely and low cost) manner. More importantly, it does not require any additional works at the data owners side like double data entry, forcing uniform data format, etc. Therefore, every party still has unlimited freedom what and how to share their own data without any intervention from outsiders that should be the fundamental spirit of internet technology.

## 5. SUMMARY AND DISCUSSION

We have introduced and explained what has been done in Indonesia to improve public access to global and local scientific data. There are two new approaches in our system, that is centralizing mirroring service to obtain significant cost reduction while the reliability is ultimately improved. Secondly mirroring also local data to make available local resources to the global communities. Although it seems a contradiction with the basic spirit of mirroring, that is to decentralize the data, our ARSIP has shown that such approaches could resolve the infrastructure lack in most developing countries.

Further advances are integrating and "collecting" local scientific data through automated but pre-configured crawling. Although the system is still in beta version, we expect this will work better than the previous one using DBRiptek. We also plan to embed the Online Calculator for Scientific Performance in ISI to perform fully automatic performance analysis for each person, institution and research project in annual basis [15]. This could be a useful tool for managements at any levels to make more transparent decision or regulation in the future.

Lastly, we would like to remark that our approaches or part of them are applicable to the developing countries, though it might be less relevant in the developed ones. We do believe that our small efforts should contribute to fill the information gap between two regions.

**Acknowledgment**

The author would like to thank the Organizer for warm hospitality during the workshop. All related works discussed in this paper fully supported by the Group for Theoretical and Computational Physics LIPI and its members. We would like to express our great acknowledgment for long collaboration with the ARSIP Team.